# Horizon-T Experiment Detailed Calibration of Cables

D. Beznosko [1a], A. Iakovlev [b], V.D. Mosunov [c], B. Mustafin [d], A. Sabyrov [d], M.I. Vildanova [c], V.V. Zhukov [c]

*Abstract*

The ability to extract the pulse width and translate it into the actual disk width of the Extensive Air Showers (EAS) is a hard one requiring accurate knowledge of the system performance. For that, the analysis for the cable calibration for Horizon-10T detectors has been re-analyzed in a different form that allows for better signal width measurements.

An innovative detector system Horizon-10T, constructed to study EAS in the energy range above $10^{16}$ eV coming from a wide range of zenith angles (0° - 85°), is located at Tien Shan high-altitude Science Station of Lebedev Physical Institute of the Russian Academy of Sciences at approximately 3340 meters above the sea level.

## 1. Detector System Description

Horizon-10T (H10T) detector system description is given in [1] [2] [3] and is now called just Horizon-T to remove any confusion. Its predecessor is Horizon-8T detector system [4] [5] and it is no longer active while the data is still being analyzed. Additionally, the short summary is shown below.

The H10T view from above is shown in Figure 1. The system center is located by a geodesic benchmark at the point 1 at the height of 3346.05 meters above the sea level and with coordinates of 43°02′49.1532" N and 76°56′43.548" E and is the origin for the coordinate system. In this XYZ system, X-axis is directed to the north, Y-axis to the west and Z-axis is directed vertically up. The geometric factor Γ of Horizon-10T is ~1.52 $km^2$/ster at $10^{17}$ eV.

Time of passage of the charged particles from EAS disk are registered at ten detection points. The relative distances of every point are presented in Table 1. All detection points have one scintillator detector (SD) in the z-plane (parallel to the sky). SD uses 5cm thick polystyrene scintillator [6] with 1 $m^2$ area (square shaped) with Hamamatsu [7] 2-inch R7723 photomultiplier tube (PMT) except points 9-10 that 12.7 cm Hamamatsu H6527 PMTs. Points 1 - 5 have fast glass-based detectors (GD) [8] [9] with R7723 PMT as a readout. Note that a previously planned upgrade to liquid scintillators [10] [11] [12] is not being considered.

[1] dbeznosko@fas.harvard.edu (also dima@dozory.us)
[a] Harvard University, Cambridge, MA, USA
[b] Computer Systems Institute, Charlestown, MA, USA
[c] P. N. Lebedev Physical Institute of the Russian Academy of Sciences, Moscow, Russia
[d] Nazarbayev Intellectual School, Astana, Kazakhstan

**Table 1: Distances of all detection points from center.**

| Point # | 1 | 2 | 3 | 4 | 5 | 6 | 7 | 8 | 9 | 10 |
|---|---|---|---|---|---|---|---|---|---|---|
| R, m | 0 | 133 | 148 | 155 | 194 | 454 | 393 | 367 | 594 | 1000 |

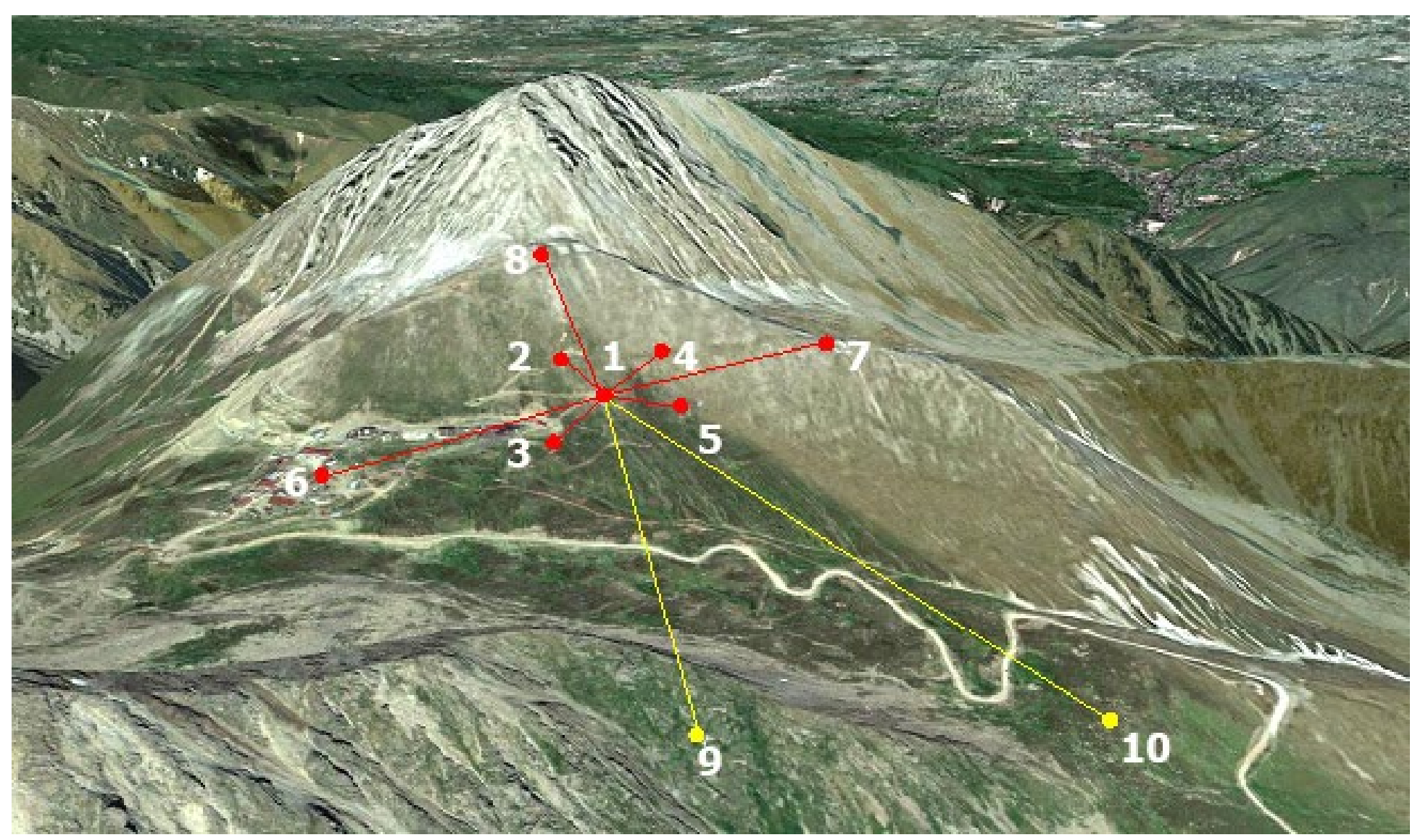

**Figure 1: H10T detection points view from above.**

### 1.1 Cables and Data Acquisition System Overview

The signals from all detectors are transmitted over coaxial cables RK 75-7-316F-C SUPER (by SpetsKabel [13]). There cables are then impedance matched to the electronics inputs and are individually calibrated [14] [15]. The master trigger is issued by a main 14-bit CAEN [16] DT5730 Analog-to-Digital converter (ADC) board. A total of three ADC boards make the data acquisition system (DAQ) and are triggered by this pulse. DAQ is positioned next to first detection point. Master trigger is issued when the detection points 3 and 1 (another option is 3 and 4) exceed detection threshold of about 4 particles. Recorded data is used for further offline analysis using different offline trigger that are less liberal.

## 2. Cables Calibration

The calibration of all cables is carried out using the pulse from a PMT. The recorded pulse and its reflection are analyzed to measure the total cable delay and pulse widening.

### 2.1 Procedure

(FEU49) from MELZ [17] (a 15cm diameter spherical-shaped cathode PMT with the spectral response from 360nm to 600nm) with a 10 cm thick circular scintillator covering the top of the cathode was used to generate the pulses for cable calibration procedure. This pulse was sent via a splitter to the ADC and a cable under testing. The other end of the cable was disconnected causing a pulse reflection that is also recorded by the ADC. The setup schematic is in Figure 2.

As described before [18], the area under the pulse is used as the measure for both the pulse width. Time between 10% and 50% of the pulse area is considered analogues to 'rise time' used in other works, and between 10% and 80% (in some older publications 90% was used) is taken as he pulse full width. The time between the 10% of the first pulse and 10% of the second pulse is taken as the double length of the cable accounting for the pulse travel along the cable and then reflecting back. Since the cables are buried, it is not possible to bring two ends together to be read by a same device for direct delay calibration.

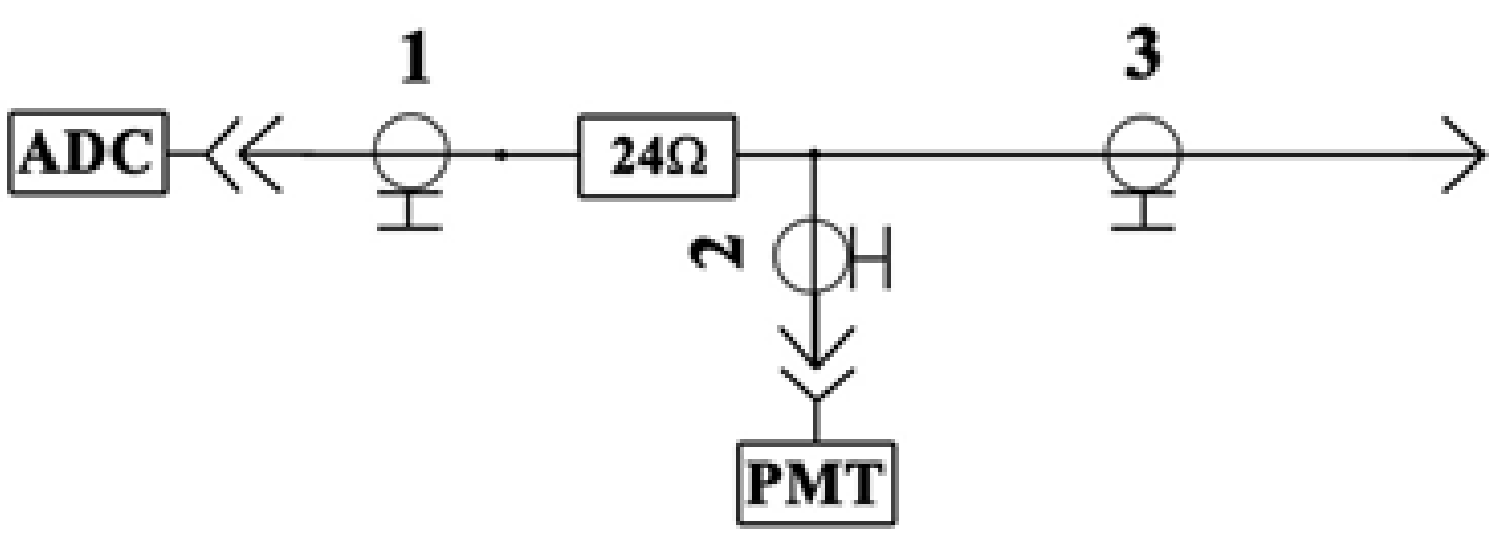

**Figure 2: Schematic of the cable calibration setup. Number 3 marks the cable under testing.**

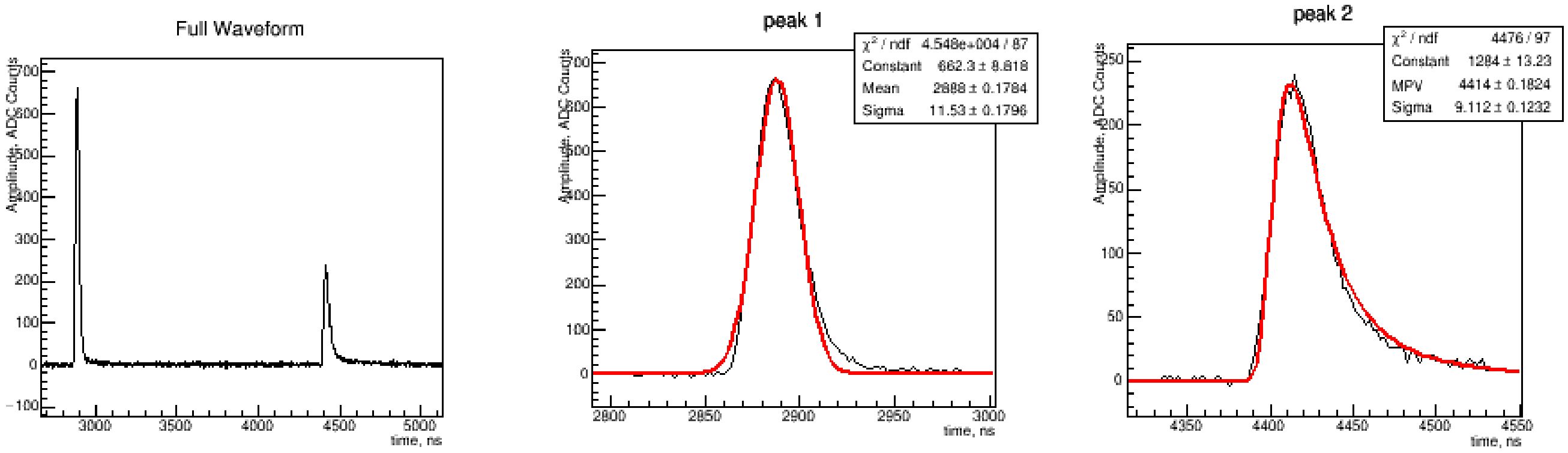

**Figure 3: Full waveform of the original + reflected pulse, a fit of both pulses.**

In Figure 3 we show the total ADC recorded waveform of the PMT pulse plus the reflection from the cable's far end. Pulse 1 is fitted with a Gaussian to accurately find its center and set an integration window. Then a simulative sum (e.g. CDF) is computed to find the times for 10%, 50% and 80% of the pulse area as shown in Figure 4 left. The right part in the same figure shows the CDF for the reflected pulse.

Note that the reflected pulse is fitted with a Landau curve in place of a Gaussian. The parameters of the fit are used to control the quality of the results for the reflected pulse as noise or high attenuation of the pulse may affect the measurement. We use that MPV- 1.09225$\sigma$ is the position of 10% of the area of the Landau curve, MPV+1.3558 $\sigma$ is 50% and MPV+5.7677 $\sigma$ is 80% (MPV+11.6493 $\sigma$ is 90% for reference).

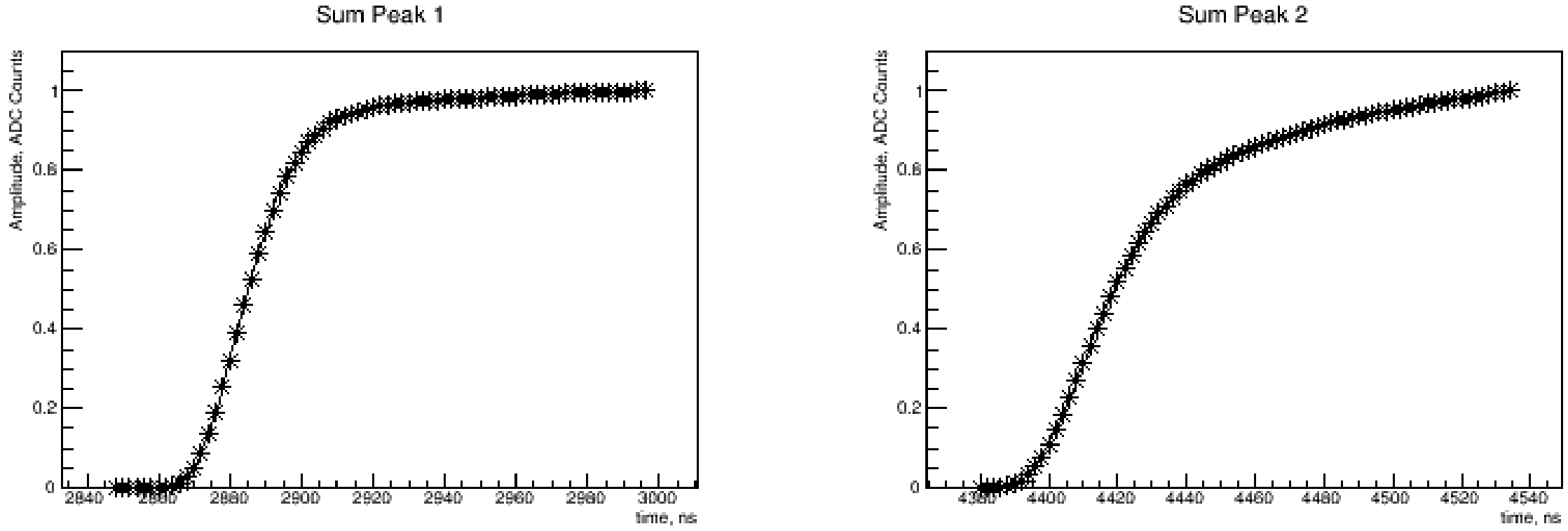

**Figure 4: Cumulative sum of the original PMT pulse and the reflected pulse.**

Each detection point has a minimum of 1 cable, some have 2 and the center one has several. That is, all points have a RK 75-7-316F-C SUPER cable, and we will refer to all such cables as ‘new’. The older cabled that initially existed in the system are calibrated separately as they are not of the same type and normally cannot be put together in a single procedure.

For all ‘new’ cables, the radio of the perfected pulse width over the original PMT width is plotted vs. the time delay between the 10% area of each pulse taken as the total distance traveled by the pulse. The plots (for 50% and 80%) are fitted by 2$^{nd}$ degree polynomial to account for the quadratic nature of the dispersion. The resulting fit is used to obtain the actual ratios for each cable using the actual cable lengths (in ns). These ratios are to be used later to remove cable widening effect from the analysis of the averaged pulse widths at the different distances from the EAS axis.

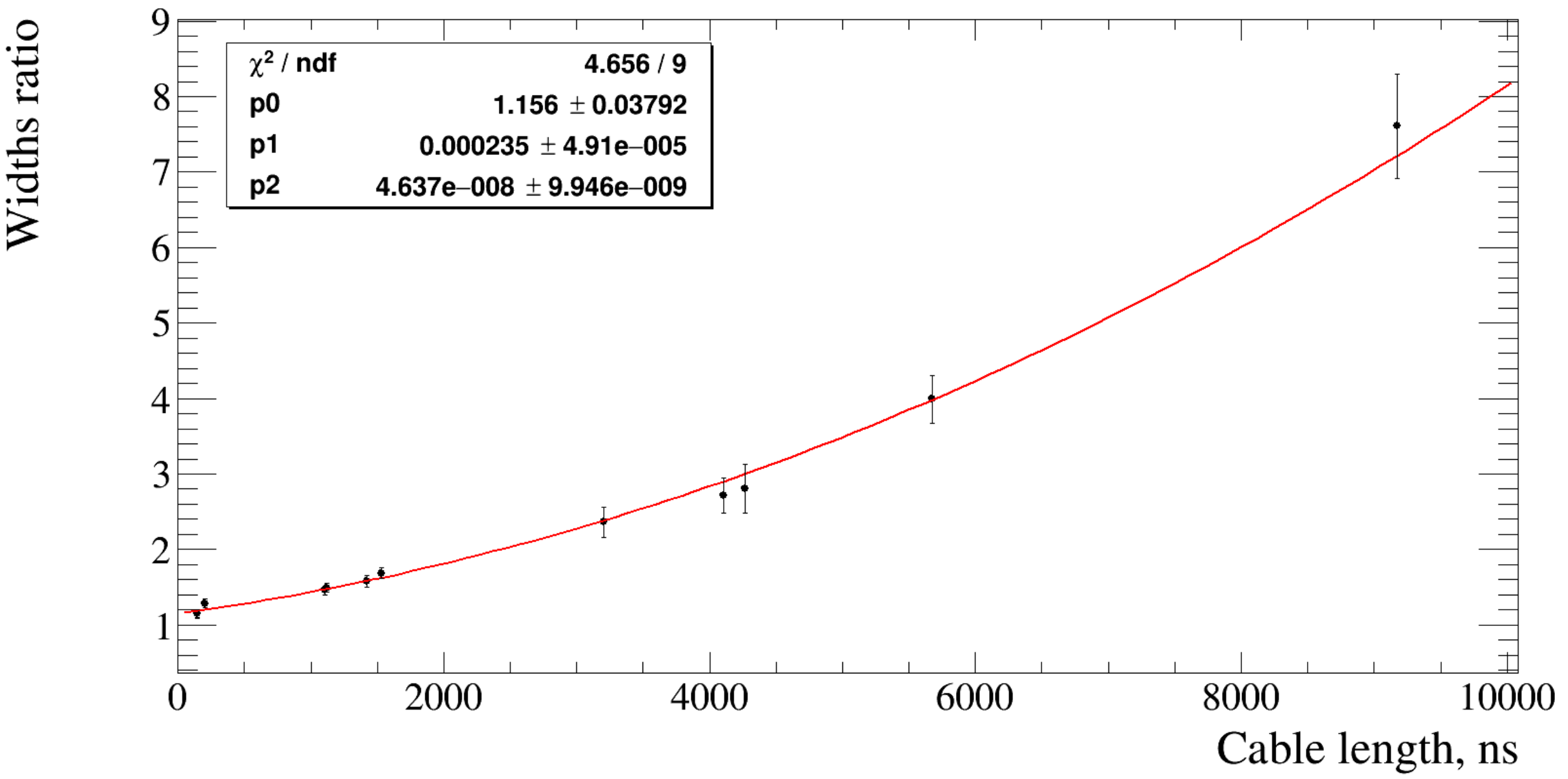

**Figure 5: Data fit for 10%-50% area pulse width ratio.**

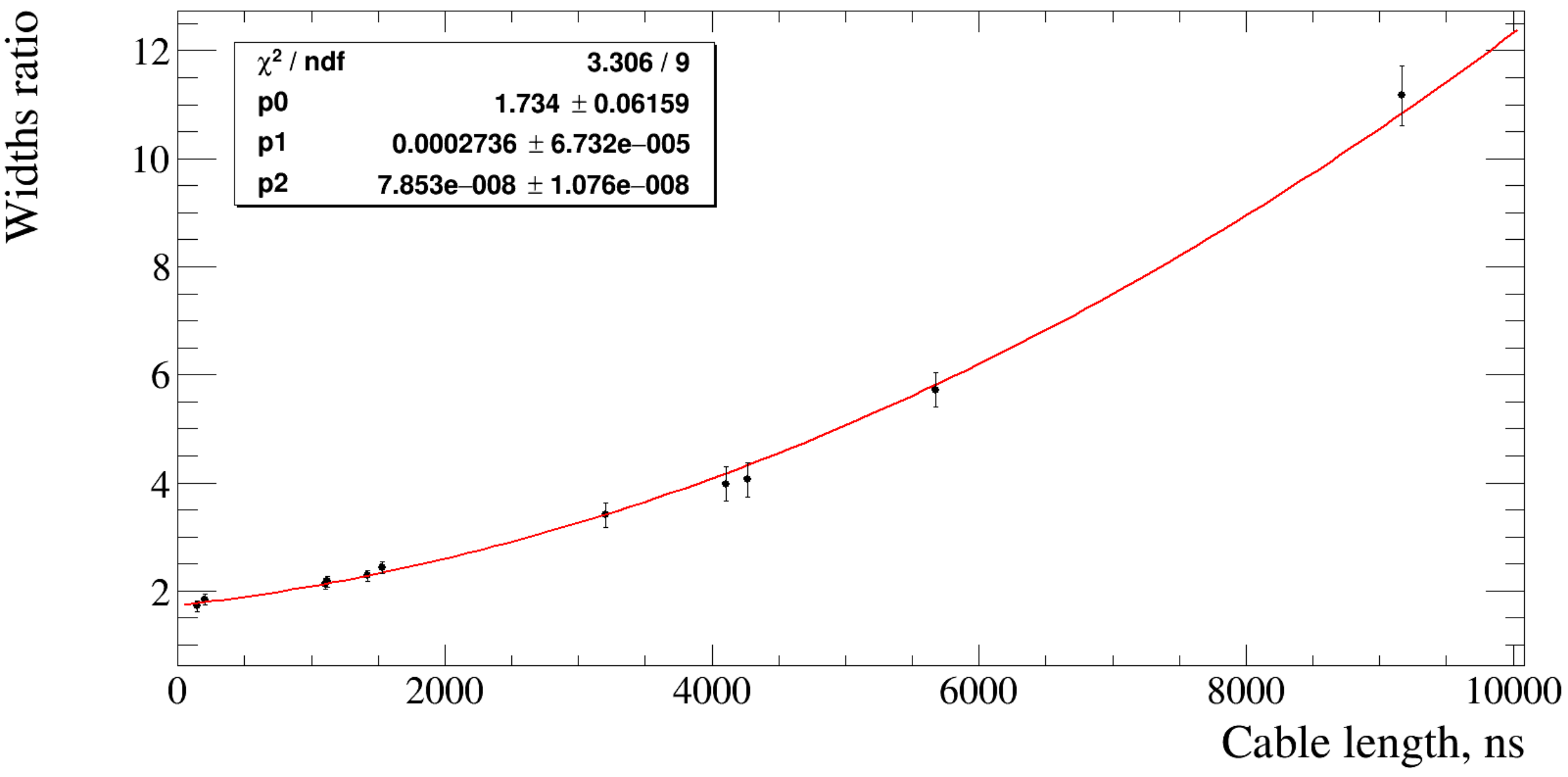

**Figure 6: Data fit for 10%-80% area pulse width ratio.**

## 2.2 Data analysis

The plots above show the reflected pulse width over the original one for the 10% to 50% area (Figure 5) and 10% to 80% (Figure 6) of each area versus the total pulse travel time for all 'new' cables. The fits to quadratic polynomial of form $p0+p1*x+p2*x^2$ is also shown in both figures.

**Table 2: Widths ratio results for all 'new' cables.**

| Cable code name | Cable length (ns) | Widths ratio for 50% | Widths ratio for 80% |
|---|---|---|---|
| Center Red-Red | 72.90 ± 0.15 | 1.17 ± 0.35 | 1.75 ± 0.44 |
| Center Yellow-Red | 72.98 ± 0.15 | 1.17 ± 0.35 | 1.75 ± 0.44 |
| Center Blue-Red H6527 | 102.35 ± 0.18 | 1.18 ± 0.35 | 1.76 ± 0.44 |
| Left New | 551.435 ± 0.22 | 1.30 ± 0.39 | 1.91 ± 0.48 |
| Right New | 561.815 ± 0.25 | 1.30 ± 0.39 | 1.91 ± 0.48 |
| Kurashkin New | 709.155 ± 0.34 | 1.35 ± 0.40 | 1.97 ± 0.49 |
| Bottom-New (green-white) | 763.91 ± 0.32 | 1.36 ± 0.41 | 1.99 ± 0.50 |
| Stone Flower New | 1601.95 ± 0.85 | 1.65 ±0.50 | 2.37 ± 0.60 |
| Upper New | 2051.22 ± 0.91 | 1.83 ± 0.55 | 2.63 ± 0.66 |
| Yastrebov New | 2132.51 ± 0.95 | 1.87 ± 0.56 | 2.67 ± 0.67 |
| 600m | 2837.11 ± 0.99 | 2.20 ± 0.66 | 3.14 ± 0.79 |
| Bunker | 4585.12 ± 3.51 | 3.21 ± 0.96 | 4.64 ± 1.16 |

We use these fits to find the ratio of pulse widths for any length of the same type of cable, in particular, actual lengths of all cables used for these measurements. The results for all ‘new’ cables are presented in Table 2. Note that the result for all central cables except one used with H6527 are the same within errors.

## 3. Conclusion

The calibration for all HT10T cables has been completed. The analysis and the results for all ‘new’ type cables is presented in this paper. Currently, H10T has the highest tine resolution and ability to resolve the pulse width as compared to other experiments such as [19] or [20].